# Enhancement of Opto-Electro-Mechanical Entanglement through Three-Level Atoms

Abebe Senbeto Kussia[a], Tewodros Yirgashewa Darge[b], Tesfay Gebremariam Tesfahannes[c,], Abeba Teklie Bimeraw[c], Berihu Teklu[d]

[a] *College of Natural Science, Department of Physics, Salale University,Fitche, 245, Ethiopia,*
[b] *College of Natural Science, Department of Applied Physics, Adama Science and Technology University, Adama, 1888, Ethiopia*
[c] *College of Natural Science, Department of Physics, Arba Minch University, Arba Minch, 21, Ethiopia*
[d] *College of Computing and Mathematical Sciences, Department of Mathematics, Khalifa University of Science and Technology, Khalifa, 127788, Abu Dhabi, United Arab Emirates*

## Abstract

We address the dynamical bipartite entanglement in an opto-electro-mechanical system that involves a three-level atom. The system consists of a degenerate three-level atom, a mechanical resonator, an optical cavity, and a microwave cavity. By utilizing the linearization approximation and nonlinear quantum-Langevin equations, the dynamics of the system are analyzed, and the bipartite entanglement is evaluated using the logarithmic negativity. The research findings indicate that the entanglement between each subsystem increases with the atom injection rate, suggesting that a higher atom injection rate leads to enhanced information transmission between the subsystems. Additionally, it is observed that the correlation between subsystems increases with an increase in the coupling rate. Moreover, the study demonstrates that the correlation between each subsystem decreases as temperature rises. We also show that the degree of tripartite entanglement diminishes with increasing atomic decay rates. The results highlight the positive impact of three-level atoms on the bipartite entanglement in an opto-electro-mechanical system. Consequently, such electro-optomechanical systems can offer a framework for optomechanical information transfer.

## 1. Introduction

Over the past few decades, cavity optomechanics, which investigates the nonlinear interaction through radiation-pressure force, has made significant progress in both experimental and theoretical domains[1, 2, 3, 4, 5]. Numerous research on optomechanical systems have been fascinating such as cooling ground state [6, 7], feedback cooling [8], normal mode splitting [9] and two-mode squeezed states [10]. Progress in those diverse optomechanical devices has been developed to perform and enhance the nonlinear quantum phenomena such as quantum entanglement [11, 12, 13, 14], high precision metrology [15, 16, 17, 18], optomechanical storage [19] and quantum information processing [20]. In quantum information processing, quantum entanglement is an important quantum mechanics phenomenon and plays a significant role in processes such as quantum teleportation [21], quantum key distribution [22], population transfer [23], quantum secure direct communication [24, 25, 26, 27] and quantum state sharing [28, 29].
Among these researches, intracavity entanglement is attainable with a coherent two-mode squeezing interaction, and steady-state entanglement is achieved in a three-mode optomechanical system [30]. Furthermore, the optomechanical dark mode successfully achieved by weakly coupling [31, 32], multimode circuit optomechanical systems may be used in the interconversion of microwave and optical photons [33, 34]. Such hybrid systems may allow to improve microwave quantum illumination [35] and to realize an information transfer between an optical and microwave cavity modes [36]. On the other hand, the nonlinear interaction of a three-level atom with a two-mode field in an optomechanical cavity enhances the non-classical properties through nonlinear atom-field coupling [37]. However, an optomechanical system with two-level and three-level atoms has been studied and has many potential applications.

The study of nano-opto-electro-mechanical systems has made remarkable strides recently, paving the way for significant advancements in this discipline and offering excellent opportunities to control light flow in nano-photonic structures [38]. Among these, opto-electro-mechanical systems have gained considerable interest due to their potential applications in theoretical and experimental frameworks. A notable example of such advancements is the work by [39], which demonstrates the generation of EPR-entangled states between optical and mechanical modes. This study showcases the dynamical entanglement and quantum steering within a pulsed hybrid opto-electro-mechanical system, highlighting the intricate interplay between different physical domains to achieve quantum control and manipulation. They propose that the mechanical oscillator performs better and more efficiently in connecting microwave and optical modes of the electromagnetic spectrum. Furthermore, Li et al. [38] have proposed an experimental demonstration of entanglement between a mechanical resonator and a microwave LC resonator in opto-electro-mechanics. This electro-

*Email address:* `tesfaye.gebremariam@amu.edu.et` (Tesfay Gebremariam Tesfahannes)

mechanical coupling serves as the source entanglement, which can exist in a stationary state and is temperature-resistant. Recent advancements in the nano-opto-electro mechanical systems offer a foundation for further research in this area and excellent opportunities to control the flow of light in nano-photonic structures [40]. Additionally, Cattiaux et al. [41] present the nonlinear dynamics of microwave optomechanics in the classical regime, suggesting that coupling optical fields to microwave signals through a mechanical resonator holds significant promise for both current and future quantum information technologies.

Moreover, the potential of the opto-electro-mechanics systems to produce the stationary entanglement dynamics between the subsystems has been investigated in [42]. In their study, the mechanical resonator simultaneously coupled to both microwave and optical cavities through a capacitor plate coupling process. Consequently, such cavity electro-mechanical systems have gained interest as potential contenders for microwave-based photonic devices. Furthermore, many schemes have been proposed to generate entanglement in nano-electro-optomechanical systems[43, 44] and in various systems[45, 46]. In addition, entanglement involving multiple modes is crucial for a wide range of quantum information and communication applications. Accordingly, particular attention has been given to entanglement between three distinct optical modes [47]. Thus, tripartite entanglement quantifiers allow to generation of the switchable genuine tripartite entanglement with Conditions necessary and sufficient for understanding the separability properties of these states [48].

Recently, Pan et al. studied the entanglement phenomena using an electro-optical hybrid system that included an optical parametric amplifier and a Coulomb force interaction. They hypothesized that the two charged oscillators improved the entanglement and output squeezing in the hybrid system. The enhancement of bipartite entanglement in nano-electro-optomechanical systems, especially in the presence of three-level atoms, remains an active area of research.
In this paper, we theoretically investigate both bipartite and tripartite entanglement dynamics within an opto-electro-mechanical system that includes a three-level atom. We examine a model in which a degenerate three-level atom is injected into the optical cavity at a constant rate. A three-level atom's upper, middle, and lower levels are labeled in a cascade configuration. The main advantage of incorporating a three-level atom into an opto-electro-mechanical system is to improve the bipartite entanglement between the subsystems under strong coupling. Moreover, the presence of the three-level atom facilitates effective state transfer and information manipulation between the subsystems. We examine the enhancement of the bipartite entanglement through logarithmic negativity. Compared to a scheme without a three-level atom, incorporating a three-level atom provides several advantages, such as enhancing the possibilities for bipartite entanglement. Our findings indicate that the entanglement between the optical cavity and the three-level atom and normalized optical cavity detuning increases with a higher coupling rate. Additionally, as the atom injection rate rises, the entanglement between each subsystem grows since a higher atom injection rate results in increased laser emission. The tripartite entanglement also analyzed using the minimal residual cotangle measure. It reveals that tripartite entanglement decreases as the atomic decay rates increase. We demonstrate that incorporating a three-level atom in the system enhances bipartite entanglement and provides insights into how atomic decay rates influence tripartite entanglement.

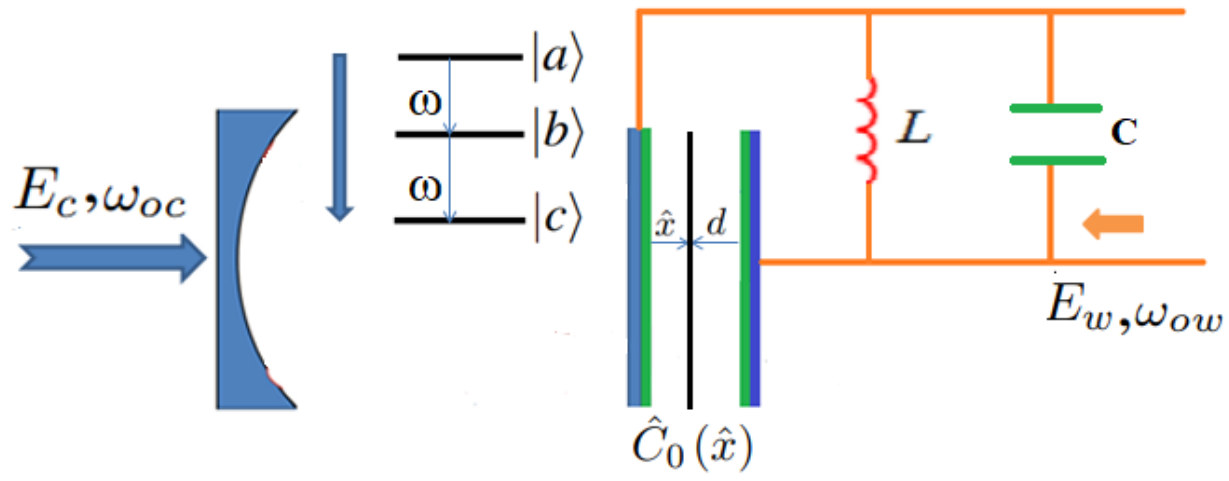

Figure 1: Schematic diagram of hybrid opto-electro-mechanical system with degenerate three-level atom.

## 2. Model and Hamiltonian Formulation

In this paper, we consider a model of opto-electro-mechanical system in which a degenerate three level atom is injected into a cavity at constant rate $r_a$ as shown in Fig. (1). In a cascade configuration, we label the top, intermediate, and bottom levels of a three-level atom as $|a\rangle$, $|b\rangle$ and $|c\rangle$, respectively. When the atom makes a transition from the upper to the intermediate level and then from the intermediate to the bottom level, two-photon modes with the same frequency are emitted [49]. We assume the transitions between levels $|a\rangle$ and $|b\rangle$ and between levels $|b\rangle$ and $|c\rangle$ to be dipole allowed, with the direct transition between levels $|a\rangle$ and $|c\rangle$ being dipole forbidden. Additionally, we assume a mechanical resonator (MR) that is linked to a driven optical cavity (OC) with a resonance frequency of $\omega_c$ and, on the other side, capacitively connected to the field of a superconducting microwave cavity (MC) of $\omega_w$.

To describe the dynamics of the system, we employ the Hamiltonian formalism. The total Hamiltonian $H$ of the system can be written as the sum of the Hamiltonians of the individual components and their interactions [50]:

$$\begin{aligned}\hat{H} = {} & \frac{\hat{p}_x^2}{2m} + \frac{m\omega_m^2\hat{x}^2}{2} + \frac{\hat{\Phi}^2}{2L} + \frac{\hat{Q}^2}{2[C+\hat{C}_o(\hat{x})]} - e(t)\hat{Q} \\ & + \hbar\omega_c\hat{a}^\dagger\hat{a} + \sum_{i=a,b,c}\hbar E_i\hat{\sigma}_{ii} - \hbar G_{oc}\hat{a}^\dagger\hat{a}\hat{x} + \hbar g(\hat{\sigma}_{ba}\hat{a}^\dagger \\ & + \hat{\sigma}_{cb}\hat{a}^\dagger + \hat{a}\hat{\sigma}_{ab} + \hat{a}\hat{\sigma}_{bc}) + i\hbar E_c(\hat{a}^\dagger e^{-i\omega_{oc}t} - \hat{a}e^{i\omega_{oc}t}),\end{aligned} \tag{1}$$

where $(\hat{x}, \hat{p}_x)$ represent the canonical position and momentum of an MR with frequency $\omega_m$ and commutation relation $[\hat{x}, \hat{p}_x] = i\hbar$, $(\hat{\Phi}, \hat{Q})$ represent the canonical coordinates of the MC with the commutation relation $[\hat{Q}, \hat{\Phi}] = i\hbar$, describing the flux through an equivalent inductor $L$ and the charge on an equivalent capacitor $C$, respectively. $G_{oc} = (\frac{\omega_c}{\ell})\sqrt{\frac{\hbar}{m\omega_m}}$ is the optomechanical coupling rate with $m$ the effective mass of mechanical mode, and $\ell$ the length of the optical Fabry-Perot cavity. The coherent driving of the microwave cavity (MC) with damping rate $\kappa_w$ is given by:

$$e(t) = -i\sqrt{2\hbar\omega_w L}E_w(e^{i\omega_{ow}t} - e^{-i\omega_{ow}t}). \tag{2}$$

The microwave resonator is driven by a tunable electric field, which acts as a control part of the system and interacts with the cavity field through the mediator of MR [51]. The microwave cavity and the MR are coupled because the microwave cavity's capacity depends on the resonator displacement, $C_o$. This function is expanded around the resonator's equilibrium position, which corresponds to the separation $d$ between the capacitor's plates and a corresponding bare capacitance of $C_o$. Expanding $\hat{C}_o(\hat{x}) = C_o[\frac{1+\hat{x}(t)}{d}]$ and using a Taylor series for the capacitive energy, we find to first order:

$$\frac{\hat{Q}^2}{2[C + \hat{C}_o(\hat{x})]} = \frac{\hat{Q}^2}{2C_\Sigma} - \frac{\mu}{2dC_\Sigma}\hat{x}(t)\hat{Q}^2, \tag{3}$$

where $C_\Sigma = C + C_o$ and $\mu = \frac{C_o}{C_\Sigma}$ [52]. Using this expression, the Hamiltonian of Eq. (1) can be rewritten in terms of the raising and lowering operators of the MC field $\hat{b}, \hat{b}^\dagger$ where $[\hat{b}, \hat{b}^\dagger] = 1$ and the dimensionless position and momentum operators of the MR, $\hat{q}, \hat{p}$ where $[\hat{q}, \hat{p}] = i\hbar$ as

$$\begin{aligned} H &= \hbar\omega_w\hat{b}^\dagger\hat{b} + \hbar\omega_c\hat{a}^\dagger\hat{a} + \sum_{i=a,b,c}\hbar E_i\hat{\sigma}_{ii} + \frac{\hbar\omega_m}{2}(\hat{p}^2 + \hat{q}^2) \\ &- \frac{\hbar G_{ow}}{2}\hat{q}(\hat{b} + \hat{b}^\dagger)^2 - \hbar G_{oc}qa^\dagger a - i\hbar E_w(e^{i\omega_{ow}t} - e^{-i\omega_{ow}t})(\hat{b} + \hat{b}^\dagger) \\ &+ i\hbar E_c(\hat{a}^\dagger e^{-i\omega_{oc}t} - \hat{a}e^{i\omega_{oc}t}) + \hbar g(\hat{\sigma}_{ba}\hat{a}^\dagger + \hat{\sigma}_{cb}\hat{a}^\dagger + \hat{a}\hat{\sigma}_{ab} + \hat{a}\hat{\sigma}_{bc}), \end{aligned} \tag{4}$$

where $\hat{b} = \frac{L\omega_w\hat{Q}+i\hat{\Phi}}{\sqrt{2L\hbar\omega_w}}$, $\hat{q} = \sqrt{\frac{m\omega_m}{\hbar}}\hat{x}$, and $G_{ow} = \frac{\mu\omega_w}{2d}\sqrt{\frac{\hbar}{m\omega_m}}$.To obtain the standard Hamiltonian of the system in terms of microwave, optical, and atom detuning, we first apply the rotating wave approximation (RWA) on Eq. (4) to eliminate fast oscillating terms [53]. The RWA assumes that the slowly varying terms of the Hamiltonian have a significant contribution to the dynamics, while the rapidly oscillating terms do not. This approximation is crucial for accurately describing the effective interactions in the system without the complexity of non-resonant terms, which have minimal effect on the system's long-term dynamics. Thus, the final Hamiltonian of the system becomes:

$$\begin{aligned} \hat{H} &= \hbar\Delta_{ow}\hat{b}^\dagger\hat{b} + \hbar\Delta_{oc}\hat{a}^\dagger\hat{a} + \hbar\Delta_{a1}\hat{\sigma}_{aa} + \hbar\Delta_{a2}\hat{\sigma}_{cc} \\ &+ \frac{\hbar\omega_m}{2}(\hat{p}^2 + \hat{q}^2) - \hbar G_{ow}\hat{q}\hat{b}^\dagger\hat{b} - \hbar G_{oc}\hat{q}\hat{a}^\dagger\hat{a} - i\hbar E_w(\hat{b} - \hat{b}^\dagger) \\ &+ i\hbar E_c(\hat{a}^\dagger - \hat{a}) + \hbar g(\hat{\sigma}_{ba}a^\dagger + \hat{\sigma}_{cb}\hat{a}^\dagger + \hat{a}\hat{\sigma}_{ab} + \hat{a}\hat{\sigma}_{bc}). \end{aligned} \tag{5}$$

where $\Delta_{ow} = \omega_w - \omega_{ow}$ is the microwave detuning with deriving frequency, $\Delta_{oc} = \omega_c - \omega_{oc}$ is optical cavity detuning, $\Delta_{a1} = \omega - \omega_{oc}$ and $\Delta_{a2} = \omega - \omega_{oc}$ are the detunings between atomic transition frequencies and driving laser frequency, with $\omega$ being the atomic transition frequencies. The last term describes the coupling between cavity mode and the three level atom with interaction strength $g$.

## 3. Dynamics of the system

The dynamics of an opto-electro-mechanical system assisted by three-level atoms can be described by a set of nonlinear quantum Langevin equations (QLEs), which incorporate both dissipation and fluctuation terms. These equations are derived from the Heisenberg equations of motion for the system operators, with additional terms representing the interaction with the environment. To obtain the nonlinear dynamics of the system operators $\hat{q}, \hat{p}, \hat{a}, \hat{b}, \hat{\sigma}_{ba}$ and $\hat{\sigma}_{cb}$, we use the Langevin-Heisenberg equation of motion. Setting $\hbar = 1$, the QLEs for the system operators can be written as follows:

$$\begin{aligned} \dot{\hat{q}} &= \omega_m\hat{p}, \\ \dot{\hat{p}} &= -\omega_m\hat{q} + \hbar G_{ow}\hat{b}^\dagger\hat{b} + \hbar G_{oc}\hat{a}^\dagger\hat{a} - \gamma_m\hat{p} + \xi, \\ \dot{\hat{a}} &= -(\kappa_c + \Delta_{oc})\hat{a} + iG_{oc}\hat{q}\hat{a} + E_c - ig(\hat{\sigma}_{ba} + \hat{\sigma}_{cb}) + \sqrt{2\kappa_c}\hat{a}^{in}, \\ \dot{\hat{b}} &= -(\kappa_w + \Delta_{ow})\hat{b} + iG_{ow}\hat{q}\hat{b} + E_w + \sqrt{2\kappa_w}\hat{b}^{in}, \\ \dot{\hat{\sigma}}_{ba} &= -(\kappa_a + i\Delta_{a1})\hat{\sigma}_{ba} + ig\hat{\sigma}_{ca}\hat{a}^\dagger - ig\hat{a}(\hat{\sigma}_{bb} - \hat{\sigma}_{aa}) + \sqrt{2\kappa_a}\hat{\sigma}^{in}_{ba}, \\ \dot{\hat{\sigma}}_{cb} &= -(\kappa_a - i\Delta_{a2})\hat{\sigma}_{cb} - ig\hat{\sigma}_{ca}\hat{a}^\dagger - ig\hat{a}(\hat{\sigma}_{cc} - \hat{\sigma}_{bb}) + \sqrt{2\kappa_a}\hat{\sigma}^{in}_{cb}, \end{aligned} \tag{6}$$

where $\gamma_m$, $\kappa_c$, $\kappa_w$ are mechanical damping rate, cavity decay rate and microwave cavity decay rate respectively. The random Brownian stochastic force $\xi(t)$ acting on the mechanical mode affects the mechanical resonator and it is coupled to a thermal bath at a damping rate of $\gamma_m$ with the correlation function [54] The optical and microwave cavity amplitude decay at a rate $\kappa_c$ and $\kappa_w$ is affected by the radiation input noise $a^{in}(t)$ and $b^{in}(t)$ whose correlation function given by [55] $\langle\hat{a}^{in}(t)\hat{a}^{\dagger in}(t')\rangle = [N(\omega_c)+1]\delta(t-t')$, $\langle\hat{a}^{\dagger in}(t)\hat{a}^{in}(t')\rangle = N(\omega_c)\delta(t-t')$, $\langle\hat{b}^{in}(t)\hat{b}^{\dagger in}(t')\rangle = [N(\omega_w) + 1]\delta(t - t')$, $\langle\hat{b}^{\dagger in}(t)\hat{b}^{in}(t')\rangle = N(\omega_w)\delta(t-t')$, where $N_{(\omega_c)} = [exp(\hbar\omega_c/\kappa_B T)-1]^{-1}$ and $N_{(\omega_w)} = [exp(\hbar\omega_w/\kappa_B T) - 1]^{-1}$ are the equilibrium mean thermal photon numbers of the optical and microwave fields respectively. One can safely assume $N_{(\omega_c)} \approx 0$ since $\hbar\omega_c/\kappa_B T \gg 1$ at optical frequencies, while thermal microwave photons cannot be neglected in general, even at very low temperatures. The correlation function of atomic input noise operators depends on the specifics of the system under consideration. In our case we assume that the correlation function of atomic input noise operators is considered as Markovian Gaussian noise, meaning it only depends on the current state and has no memory of past states which is described by delta function as $\langle\hat{\sigma}^{in}_{ba}(t)\hat{\sigma}^{in}_{ab}(t')\rangle = [N(\omega)+1]\delta(t-t')$, $\langle\hat{\sigma}^{in}_{ab}(t)\hat{\sigma}^{in}_{ba}(t')\rangle = N(\omega)\delta(t-t')$, $\langle\hat{\sigma}^{in}_{cb}(t)\hat{\sigma}^{in}_{bc}(t')\rangle = [N(\omega) + 1]\delta(t - t')$, $\langle\hat{\sigma}^{in}_{bc}(t)\hat{\sigma}^{in}_{cb}(t')\rangle = N(\omega)\delta(t - t')$, where $N_{(\omega)} = [exp(\hbar\omega/\kappa_B T) - 1]^{-1}$ and $N_{(\omega)} = [exp(\hbar\omega/\kappa_B T) - 1]^{-1}$. We assuming that the initial state of a three-level atom $\rho_a = \rho^0_{aa}|a\rangle\langle a| + \rho^0_{cc}|c\rangle\langle c| + \rho^0_{ca}(|c\rangle\langle a| + |a\rangle\langle c|)$ to be injected in to the cavity at constant rate of $r_a$ [56]. As the result we can rewrite the last two eqauations of Eq. (6) as

$$\dot{\hat{\sigma}}_{ba} = -(\kappa_a + i\Delta_{a1})\hat{\sigma}_{ba} + igr_a\rho^0_{ca}\hat{a}^\dagger + igr_a\rho^0_{aa}\hat{a} + \sqrt{2\kappa_a}\hat{\sigma}^{in}_{ba}, \quad (7)$$

$$\dot{\hat{\sigma}}_{cb} = -(\kappa_a - i\Delta_{a2})\hat{\sigma}_{cb} - igr_a\rho^0_{ca}\hat{a}^\dagger - igr_a\rho^0_{cc}\hat{a} + \sqrt{2\kappa_a}\hat{\sigma}^{in}_{cb}. \quad (8)$$

The dynamics of open quantum systems in interaction with a reservoir, which is typically modeled as a collection of harmonic oscillators that are linearly coupled to the system. The linearization of the above nonlinear Quantum langevin equations of the system operators is done by assuming that the light mode is strongly driven leading to the amplitude of its mean value $|\alpha_s| \gg 1$. This allows us to linearize the dynamics of our system around the steady-state values by expanding our system operators $\hat{q}$, $\hat{p}$, $\hat{a}$, $\hat{b}$, $\hat{\sigma}_{ba}$, and $\hat{\sigma}_{cb}$ as a sum of its coherent amplitudes semi-classical steady-state value plus an additional small fluctuation operator with zero mean values [57].

### 3.1. Steady State Solutions of the Equations of Motion

The steady-state solution of the system refers to finding a stable equilibrium condition in which the opto-electro-mechanical system exhibits constant behavior over time, with no net changes in dynamic parameters. The steady-state solution is critical for understanding the system's long-term behavior, stability, and performance. To obtain the steady-state solution of the system operators, we set the time derivatives to zero and calculate the mean values of the operators. The steady-state does not fluctuate with time. Using the Langevin equations from Eqs.(6) and Eqs.(7)-(8), and noting that the mean values of the input noise terms and fluctuation terms are zero, the steady-state solutions of the system operators become:

$$\begin{aligned} p_s = 0, \quad q_s &= \frac{G_{oc}|\alpha_s|^2 + G_{ow}|\beta_s|^2}{\omega_m}, \\ \alpha_s &= \frac{E_c - ig(\sigma_{ba_s} + \sigma_{cb_s})}{i\Delta_c + \kappa_c}, \quad \beta_s = \frac{E_w}{i\Delta_w + \kappa_w}, \\ \sigma_{ba_s} &= \frac{igr_a(\rho^0_{ca} + \rho^0_{aa})}{\kappa_a + i\Delta_{a1}}\alpha_s, \quad \sigma_{cb_s} = -\frac{igr_a(\rho^0_{ca} + \rho^0_{cc})}{\kappa_a - i\Delta_{a2}}\alpha_s, \end{aligned} \quad (9)$$

where $\Delta_c = \Delta_{oc} - G_{oc}q_s$ and $\Delta_w = \Delta_{ow} - G_{ow}q_s$ describe the effective detuning of the optical and microwave cavities field, respectively. The dynamics of the fluctuation of the operators in the system are obtained by decomposing each operator as a sum of its steady-state value and a small fluctuation. The fluctuation of the operators of the system is obtained by Eqs. (6) and Eqs.(7)-(8) and neglect the second order fluctuations (small terms), $\delta\hat{q}\delta\hat{a} \approx \delta\hat{q}\delta\hat{b} \approx \delta\hat{q}\delta\hat{a}^\dagger \approx \delta\hat{q}\delta\hat{b}^\dagger \approx 0$. The fluctuation of the system operators can be written as:

$$\begin{aligned} \delta\dot{\hat{q}} &= \omega_m\delta\hat{p}, \\ \delta\dot{\hat{p}} &= -\omega_m\delta\hat{q} - \gamma_m\delta\hat{p} + G_{oc}\alpha_s(\delta\hat{a}^\dagger + \delta\hat{a}) + G_{ow}\beta_s(\delta\hat{b}^\dagger + \delta\hat{b}) + \xi, \\ \delta\dot{\hat{a}} &= -(i\Delta_c + \kappa_c)\delta\hat{a} + iG_{oc}\alpha_s\delta\hat{q} - ig(\delta\hat{\sigma}_{ba} + \delta\hat{\sigma}_{cb}) + \sqrt{2\kappa_c}\hat{a}^{in}, \\ \delta\dot{\hat{b}} &= -(i\Delta_w + \kappa_w)\delta\hat{b} + iG_{ow}\beta_s\delta\hat{q} + \sqrt{2\kappa_w}b^{in}, \\ \dot{\hat{\sigma}}_{ba} &= -(\kappa_a + i\Delta_{a1})\hat{\sigma}_{ba} + igr_a\rho^0_{ca}\hat{a}^\dagger + igr_a\rho^0_{aa}\hat{a} + \sqrt{2\kappa_a}\hat{\sigma}^{in}_{ba}, \\ \dot{\hat{\sigma}}_{cb} &= -(\kappa_a - i\Delta_{a2})\hat{\sigma}_{cb} - igr_a\rho^0_{ca}\hat{a}^\dagger - igr_a\rho^0_{cc}\hat{a} + \sqrt{2\kappa_a}\hat{\sigma}^{in}_{cb}, \end{aligned} \quad (10)$$

where $\alpha_s$ and $\beta_s$ are real and positive in the chosen phase reference for the optical and microwave field.

### 3.2. The quantum fluctuations and covariance matrix of the system

To study the entanglement properties of opto-electro-mechanical systems, it is convenient to define the quantum correlations between appropriate quadratures of the two intracavity fields and the position and momentum of the resonator. Hence, we introduce the dimensionless quadrature fluctuation for each system operator as $\delta\hat{X}_c = \frac{(\delta\hat{a}+\delta\hat{a}^\dagger)}{\sqrt{2}}, \delta\hat{Y}_c = \frac{(\delta\hat{a}-\delta\hat{a}^\dagger)}{i\sqrt{2}}, \delta\hat{X}_w = \frac{(\delta\hat{b}+\delta\hat{b}^\dagger)}{\sqrt{2}}, \delta\hat{Y}_w = \frac{(\delta\hat{b}-\delta\hat{b}^\dagger)}{i\sqrt{2}}$, and the microwave cavity field fluctuation quadratures as $\delta\hat{X}_{a_1} = \frac{(\delta\hat{\sigma}_{ba}+\delta\hat{\sigma}_{ab})}{\sqrt{2}}, \delta\hat{Y}_{a_1} = \frac{(\delta\hat{\sigma}_{ba}-\delta\hat{\sigma}_{ab})}{i\sqrt{2}}, \delta\hat{X}_{a_2} = \frac{(\delta\hat{\sigma}_{cb}+\delta\hat{\sigma}_{bc})}{\sqrt{2}}$, $\delta\hat{Y}_{a_2} = \frac{(\delta\hat{\sigma}_{cb}-\delta\hat{\sigma}_{bc})}{i\sqrt{2}}$. Using the above equations, the linearized QLEs describing the system quadrature fluctuations for our proposed model becomes

$$\begin{aligned} \delta\dot{\hat{q}} &= \omega_m\delta\hat{p}, \\ \delta\dot{\hat{p}} &= -\omega_m\delta\hat{q} - \gamma_m\delta\hat{p} + G_c\delta\hat{X}_c + G_w\delta\hat{X}_w + \xi, \\ \delta\dot{\hat{X}}_c &= -\kappa_c\delta\hat{X}_c + \Delta_c\delta\hat{Y}_c + g\delta\hat{Y}_{a_1} + \delta\hat{Y}_{a_2} + \sqrt{2\kappa_c}\hat{X}^{in}_c, \\ \delta\dot{\hat{Y}}_c &= G_c\delta\hat{q} - \Delta_c\delta\hat{X}_c - \kappa_c\delta\hat{Y}_c - g\delta\hat{X}_{a_1} - g\delta\hat{X}_{a_2} + \sqrt{2\kappa_c}\hat{Y}^{in}_c, \\ \delta\dot{\hat{X}}_w &= -\kappa_w\delta\hat{X}_w + \Delta_w\delta\hat{Y}_w + \sqrt{2\kappa_w}\hat{X}^{in}_w, \\ \delta\dot{\hat{Y}}_w &= G_w\delta\hat{q} - \Delta_w\delta\hat{X}_w - \kappa_w\delta\hat{Y}_w + \sqrt{2\kappa_w}\hat{Y}^{in}_w, \\ \delta\dot{\hat{X}}_{a_1} &= gr_a(\rho^0_{ca} - \rho^0_{aa})\delta\hat{Y}_c - \kappa_a\delta\hat{X}_{a_1} + \Delta_{a1}\delta\hat{Y}_{a_1} + \sqrt{2\kappa_a}\hat{X}^{in}_{a_1}, \\ \delta\dot{\hat{Y}}_{a_1} &= gr_a(\rho^0_{ca} + \rho^0_{aa})\delta\hat{X}_c - \Delta_{a1}\delta\hat{X}_{a_1} - \kappa_a\delta\hat{Y}_{a_1} + \sqrt{2\kappa_a}\hat{Y}^{in}_{a_1}, \\ \delta\dot{\hat{X}}_{a_2} &= gr_a(\rho^0_{cc} - \rho^0_{ca})\delta\hat{Y}_c - \kappa_a\delta\hat{X}_{a_2} - \Delta_{a2}\delta\hat{Y}_{a_2} + \sqrt{2\kappa_a}\hat{X}^{in}_{a_2}, \\ \delta\dot{\hat{Y}}_{a_2} &= gr_a(\rho^0_{ca} + \rho^0_{cc})\delta\hat{X}_c + \Delta_{a2}\delta\hat{X}_{a_2} - \kappa_a\delta\hat{Y}_{a_2} + \sqrt{2\kappa_a}\hat{Y}^{in}_{a_2}, \end{aligned} \quad (11)$$

where $G_c = \sqrt{2}G_{oc}\alpha_s = \frac{\omega_c}{\ell}\sqrt{\frac{2\hbar}{m\omega_m}}\left[\frac{E_c - ig(\sigma_{ba_s}+\sigma_{cb_s})}{\kappa_c + i\Delta_c}\right]$ and $G_w = \sqrt{2}G_{ow}\beta_s = \frac{\mu\omega_w}{d}\sqrt{\frac{P_w\kappa_w}{m\omega_m\omega_{ow}(\kappa_w^2+\Delta_w^2)}}$. Once we get the expressions for these fluctuation operators (Eq. (11)), we can write a matrix equation of the form [58]

$$\dot{u}(t) = Au(t) + n(t), \quad (12)$$

In which $u(t) = (\delta\hat{q}(t), \delta\hat{p}(t), \delta\hat{X}_c(t), \delta\hat{Y}_c(t), \delta\hat{X}_w(t), \delta\hat{Y}_w(t), \delta\hat{X}_{a_1}(t), \delta\hat{Y}_{a_1}(t), \delta\hat{X}_{a_2}(t), \delta\hat{Y}_{a_2})^T$, is quadrature fluctuation vector, $n(t) = (0, \xi(t), \sqrt{2\kappa}\hat{X}^{in}_c(t), \sqrt{2\kappa}\hat{Y}^{in}_c(t), \sqrt{2\kappa}\hat{X}^{in}_w(t), \sqrt{2\kappa}\hat{Y}^{in}_w(t), \sqrt{2\kappa_a}\hat{X}^{in}_{a_1}, \sqrt{2\kappa_a}\hat{Y}^{in}_{a_1}, \sqrt{2\kappa_a}\hat{X}^{in}_{a_2}, \sqrt{2\kappa_a}\hat{Y}^{in}_{a_2})^T$ is the noise vector with the exponent $T$ represent the transpose.

The steady-state dynamics of the covariance matrix for an opto-electro-mechanical system are concerned with understanding how the statistical properties of the system reach equilibrium over time. Analyzing the covariance matrix provides insights into the correlations and entanglement between different parts of the OMEMS, which are crucial for understanding its quantum behavior, performance, and potential applications. Eq. (12) is a covariance matrix form. The drift matrix $A$ can be expressed as follows:

$$A=\begin{pmatrix} 0 & \omega_m & 0 & 0 & 0 & 0 & 0 & 0 & 0 & 0 \\ -\omega_m & -\gamma_m & G_c & 0 & G_w & 0 & 0 & 0 & 0 & 0 \\ 0 & 0 & -\kappa_c & \Delta_c & 0 & 0 & 0 & g & 0 & g \\ G_c & 0 & -\Delta_c & -\kappa_c & 0 & 0 & -g & 0 & -g & 0 \\ 0 & 0 & 0 & 0 & -\kappa_w & \Delta_w & 0 & 0 & 0 & 0 \\ G_w & 0 & 0 & 0 & -\Delta_w & -\kappa_w & 0 & 0 & 0 & 0 \\ 0 & 0 & 0 & gr_a(\rho^0_{ca}-\rho^0_{aa}) & 0 & 0 & -\kappa_a & \Delta_{a1} & 0 & 0 \\ 0 & 0 & gr_a(\rho^0_{ca}+\rho^0_{aa}) & 0 & 0 & 0 & -\Delta_{a1} & -\kappa_a & 0 & 0 \\ 0 & 0 & 0 & gr_a(\rho^0_{cc}-\rho^0_{ca}) & 0 & 0 & 0 & 0 & -\kappa_a & -\Delta_{a2} \\ 0 & 0 & gr_a(\rho^0_{cc}+\rho^0_{ca}) & 0 & 0 & 0 & 0 & 0 & \Delta_{a2} & -\kappa_a \end{pmatrix}. \quad (13)$$

Due to the linearized dynamics and the Gaussian nature of the quantum noises in Eq. (12), the steady state of the quantum fluctuations of the system is a continuous variable bipartite Gaussian state, which is completely characterized by the $10\times 10$ correlation matrix $V$ with its entries are defined as [59]: $V_{ij}=\frac{1}{2}\langle u_i(t)u_j(t')+u_j(t')u_i(t)\rangle$, where $(i,j=1,2,...,10)$. Eq. (12) can be written in compact form as $\dot{u}(t)=Au(t)+n(t)$ whose solution is

$$u_i(t)=M(t)u(0)+\int_0^t dt' M(t')n_k(t-t'), \quad (14)$$

where $M(t)=exp(At)$. According to the Routh-Hurwitz criterion [60], the system reaches around a steady state and stable. i.e.; when all the eigenvalues of the coefficient matrix $A$ have negative real parts. As $t\rightarrow\infty$ so that $M(\infty)=0$ and Eq. (14) can be rewritten as

$$u_i(\infty)=\int_0^\infty dt' \sum_k M(t')n_k(t-t'). \quad (15)$$

Consequently using Eq. (15) we can write $V_{ij}=\frac{1}{2}\langle u_i(\infty)u_j(\infty)+u_j(\infty)u_i(\infty)\rangle$. By using the steady-state solutions in Eqs. (14 - 15) and the fact that the ten components of $n(t)$ are uncorrelated with each other, one gets

$$V_{ij}=\sum_{l,k}\int_0^\infty dt\int_0^\infty dt' M_{ik}(t)M_{jl}(t')n_{kl}(t-t'), \quad (16)$$

where $n_{kl}(t-t')=\frac{1}{2}\langle n_k(t)n_l(t')+n_l(t')u_k(t)\rangle$ is the matrix of stationary noise correlation functions. The Brownian noise term $\xi(t)$ is, in general, a non-Markovian Gaussian noise, but in the limit of large mechanical quality factor $Q_m=\omega_m/\gamma_m\gg 1$, becomes with a good approximation Markovian [61] with the symmetrized correlation function $\langle\xi(t)\xi(t')+\xi(t')\xi(t)\rangle = 2\gamma_m(2n+1)\delta(t-t')$, where $n=[exp(\frac{\hbar\omega_m}{\kappa_B T})-1]^{-1}$ is the mean thermal excitation number of the resonator. As a consequence, $n_{kl}(t-t')=D_{kl}\delta(t-t')$, where $D_{kl}=Diag[0,\gamma_m(2n+1),\kappa_c,\kappa_c,\kappa_w(2N(\omega_w)+1),\kappa_w(2N(\omega_w)+1),\kappa_a,\kappa_a,\kappa_a,\kappa_a]$ is the diffusion matrix stemming from the noise correlations. So that Eq. (16) becomes

$$V_{ij}=\int_0^\infty d_s M_{(s)}DM^T_{(s)}. \quad (17)$$

When the stability condition is fulfilled, the steady-state CM in Eq. (17) is determined by the Lyapunov equation [62]

$$AV+VA^T=-D, \quad (18)$$

which is linear in V and can be straightforwardly solved. However, its explicit solution is cumbersome and will not be reported here. We have a five-mode Gaussian state characterized by covariance matrix $V$, which can also be expressed in the form of a block matrix as:

$$V=\begin{pmatrix} V_m & V_{ma} & V_{mb} & V_{m\sigma_{ba}} & V_{m\sigma_{cb}} \\ V^T_{ma} & V_a & V_{ab} & V_{a\sigma_{ba}} & V_{a\sigma_{cb}} \\ V^T_{mb} & V^T_{ab} & V_b & V_{b\sigma_{ba}} & V_{b\sigma_{cb}} \\ V^T_{m\sigma_{ba}} & V^T_{a\sigma_{ba}} & V^T_{b\sigma_{ba}} & V_{\sigma_{ba}} & V_{\sigma_{ba}\sigma_{cb}} \\ V^T_{m\sigma_{cb}} & V^T_{a\sigma_{cb}} & V^T_{b\sigma_{cb}} & V^T_{\sigma_{cb}\sigma_{cb}} & V_{\sigma_{cb}} \end{pmatrix}, \quad (19)$$

where each block represents $2\times 2$ matrix. The diagonal blocks show variance within each subsystem, while the off-diagonal blocks show covariance across different subsystems [63].

## 4. Opto-electro-mechanical Entanglemnt Generations

We are interested in the entanglement properties of the steady state of the opto-electro-mechanical system under study. Therefore, we shall focus on the entanglement of the three possible bipartite subsystems. The sub-matrix representing the covariance of the cavity and mechanical subsystem is determined by the first four rows and columns of V. Accordingly, the correlation between an optical cavity-mechanical oscillator ($V_s$), microwave cavity-mechanical oscillator ($V_r$), optical cavity-microwave cavity ($V_z$) subsystem can be represented by:

$$V_s=\begin{pmatrix} V_m & V_{ma} \\ V^T_{ma} & V_a \end{pmatrix}, V_r=\begin{pmatrix} V_m & V_{mb} \\ V^T_{mb} & V_b \end{pmatrix}, V_z=\begin{pmatrix} V_a & V_{ab} \\ V^T_{ab} & V_b \end{pmatrix}, \quad (20)$$

where the index $m$, $a$ and $b$ represents MR, OC and MC subsystem respectively. The correlation between optical cavity-atom operators are

$$V_t=\begin{pmatrix} V_a & V_{a\sigma_{ba}} \\ V^T_{a\sigma_{ba}} & V_{\sigma_{ba}} \end{pmatrix}, V_y=\begin{pmatrix} V_a & V_{a\sigma_{cb}} \\ V^T_{a\sigma_{cb}} & V_{\sigma_{cb}} \end{pmatrix}, \quad (21)$$

where the index $\sigma_{ba}$ and $\sigma_{cb}$ represents atomic operators subsystem. To quantify bipartite entanglement among different subsystems, we use logarithmic negativity [64, 65]

$$E_N = max[0, -ln2\eta^-], \tag{22}$$

where $\eta^- = 2^{-\frac{1}{2}}\left[\Sigma(V_k) - \sqrt{\Sigma(V_k)^2 - 4detV_k}\right]^2$ is the lowest symplectic eigenvalue of the partial transpose of the $4 \times 4$ CM, associated with the selected bipartition, obtained by neglecting the rows and columns of the uninteresting mode, where $V_k$ $(k = s, r, z, t, y)$ can be expressed in a form of block matrix as Eq.(20) and Eq. (21).

In addition, we investigate tripartite entanglement between the modes of our system. To effectively describe tripartite entanglement, we employ the minimal residual cotangle measure [66]

$$R_{i|jk} = C_{i|jk} - C_{i|j} - C_{i|k}. \tag{23}$$

The above measure evaluates the cotangle $C_{x|y}$ between subsystems $x$ and $y$. The cotangle $C_{x|y}$ is further defined by its squared logarithmic negativity, $E^2_{x|y}$. For one-mode versus one-mode entanglement, the logarithmic negativity is defined as shown in Eq.(22). Conversely, for one-mode versus two-mode entanglement, the logarithmic negativity is defined as

$$E_{i|jk} = \max[0, -\ln(2v_{i|jk})], \tag{24}$$

where $v_{i|jk} = \min|\bigoplus_{r=1}^{3} i\sigma_y p_{i|jk} V p_{i|jk}|$. Here, matrices for partial transposition at the level of covariance matrix (CM) are denoted by:$p_{m|ab} = \sigma_z \bigoplus 1 \bigoplus 1$, $p_{a|mb} = 1 \bigoplus \sigma_z \bigoplus 1$, and $p_{a|mb} = 1 \bigoplus 1 \bigoplus \sigma_z$. Therefore, we can express the tripartite entanglement of the system as follows:

$$R_{min} \equiv \min[R_{m|ab}, R_{m|a}, R_{m|b}]. \tag{25}$$

The numerical results of the analysis of tripartite entanglement between subsystems of an opto-electro-mechanical system assisted by a three-level laser. Specifically, the entanglement properties can be verified by experimentally measuring the corresponding CM through logarithmic negativity. To measure $E_N$ at the steady state, one has to measure all the determined independent entries of the correlation matrix V. We have taken the parameters analogous to the values of the parameter used in experimental work from [56, 67, 68] listed as follow.

Furthermore, the optical cavity length $L = 1 \times 10^{-3}m$, the spacing of capacitor plate $d = 100 \times 10^{-9}m$, permeability $\mu = 0.008$, mechanical resonator mass $m = 10 \times 10^{-12}Kg$, temperature $T = 15 \times 10^{-3}K$, mechanical resonator frequency $\omega_m = 2\pi \times 10^7 Hz$, Microwave cavity frequency $\omega_\omega = 2\pi \times 10^7 Hz$, mechanical quality factor $Q = 5 \times 10^4$, mechanical decay rate $\gamma_m = \omega_m/Q$ with $F = 4.07 \times 10^4$, optical cavity decay rate $\kappa_c = 0.08\omega_m$, microwave cavity decay rate $\kappa_\omega = 0.02\omega_m$, optical cavity detuning $\Delta_w = \omega_m$, optical cavity beam wavelength $\lambda_{oc} = 810 \times 10^{-9}m$, optical cavity beam frequency $\omega_{oc} = 2\pi c/\lambda_{oc}$, optical cavity driven power $P_c = 30 \times 10^{-3}W$,

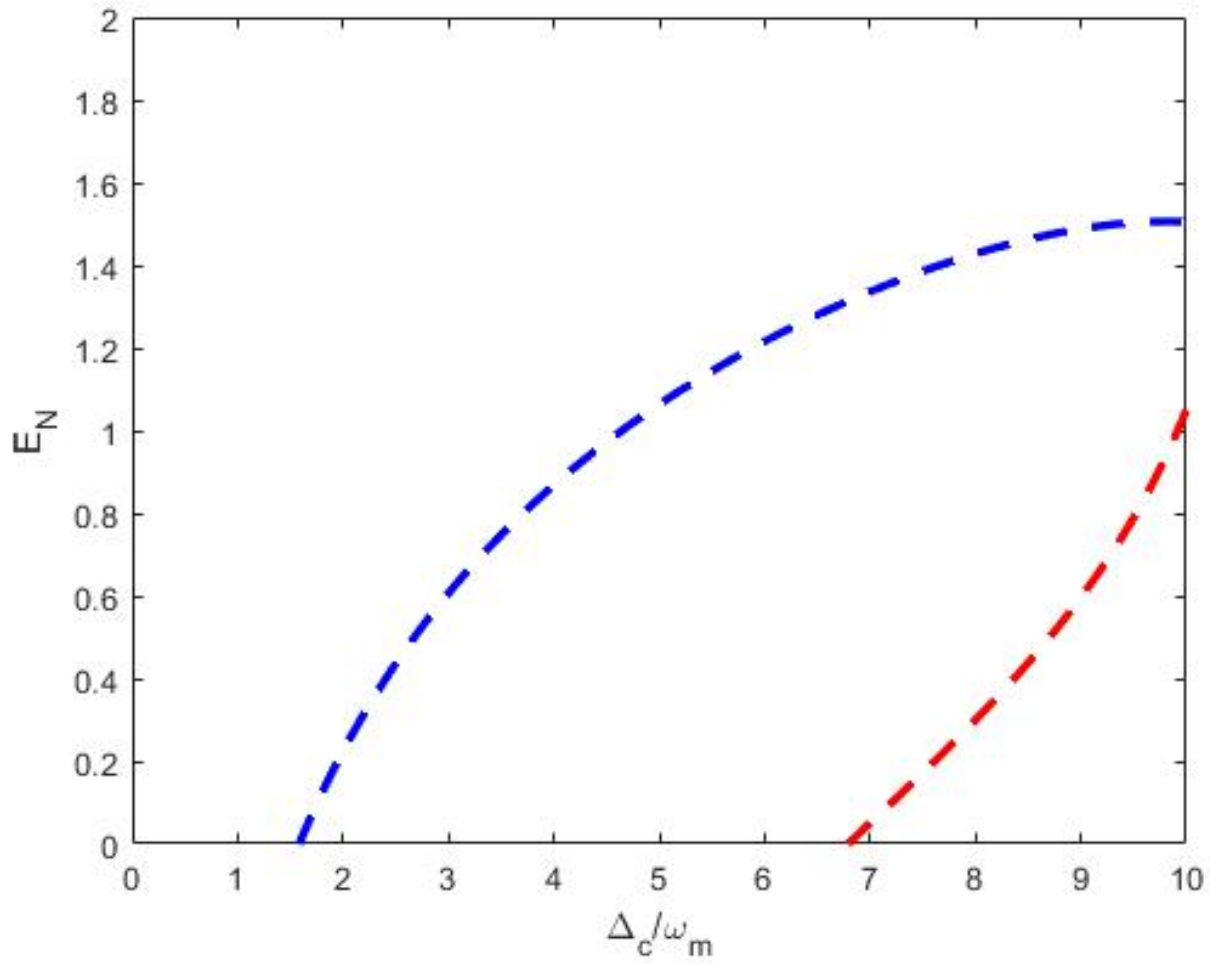

Figure 2: The plot of logarithmic negativity $E_N$ between OC-MR versus normalize optical cavity detuning $\Delta_c/\omega_m$, with out injection of atom (red line) and with injection of atom (blue line), the parameters we used are $L = 1 \times 10^{-3}m$, $d = 100 \times 10^{-9}m$, $\mu = 0.008$, $m = 10 \times 10^{-12}Kg$, $T = 15 \times 10^{-3}K$, $\omega_m = 2\pi \times 10^7 Hz$, $\omega_\omega = 2\pi \times 10^7 Hz$, $\gamma_m = 200\pi$, $\kappa_c = 0.1\omega_m$, $\kappa_w = 0.08\omega_m$, $\Delta_w = \omega_m$, $\lambda_{oc} = 810 \times 10^{-9}m$, $\omega_{oc} = 2\pi c/\lambda_{oc}$, $P_c = 30 \times 10^{-3}W$, $P_\omega = 30 \times 10^{-3}W$, $r_a = 1.6 \times 10^5$, $g = 2\pi \times 8 \times 10^5$, $\Delta_{a1} = 2\pi \times 10^{10}$, $\Delta_{a2} = 2\pi \times 10^7$, $\rho^0_{aa} = \rho^0_{cc} = \rho^0_{ca} = 0.5$, the other parametres are listed in main text.

the microwave cavity driven power $P_\omega = 30 \times 10^{-3}W$, and internal energy of the atoms are assumed to be $\rho^0_{aa} = \rho^0_{cc} = \rho^0_{ca} = 0.5$, $\Delta_{a1} = \Delta_{a2} = 2\pi \times 10^7$ are atom transitions detuning respectively

Accordingly, we study the effects of injected atoms on the entanglement between subsystems of the opto-electro-mechanical system and compare it with the standard opto-electromechanical system without three-level of atoms. To this aim, Fig. (2) shows that the plot of logarithmic negativity $E_N$ between OC-MR versus normalized optical cavity detuning $\Delta_c/\omega_m$. The red dashed line illustrates the entanglement level within the original OC-MR, whereas the blue dashed line presents the entanglement level of the OC-MR enhanced by a three-level laser. By introducing three-level atoms into the opto-electro-mechanical system, higher degrees of entanglement can be achieved due to the added three-level laser-field interaction. In this case, the entanglement between the optical and mechanical components can be improved by the atomic degree of freedom. This enhancement enables more robust and efficient entanglement generation and more control over the entangled state.

In Fig. (3), we can see two different entanglement scenarios regarding the interaction between an MC-MR, one without the injection of a three-level atom (red dashed line) and one where a three-level atom is injected into the cavity (blue dashed line). In the first scenario (red dashed line), the entanglement between MC-MR is generated solely from the interaction between the two subsystems without atom injection. This curve thus depicts the baseline level of entanglement generated by the interaction of the two subsystems. The entanglement arises due to the radiation pressure of the optical cavity on the MR, which results in a change in the length of the capacitor connecting the two sub-systems. On the other hand, the blue dashed line represents the entanglement between the MC-MR when a three-level atom is injected into the optical cavity at the rate $r_a$. By comparing the red and blue dashed lines, one can observe that the presence of a three-level laser indeed enhances the entanglement between the MC-MR, especially for specific values of the normalized detuning parameter, $\Delta_c/\omega_m$.

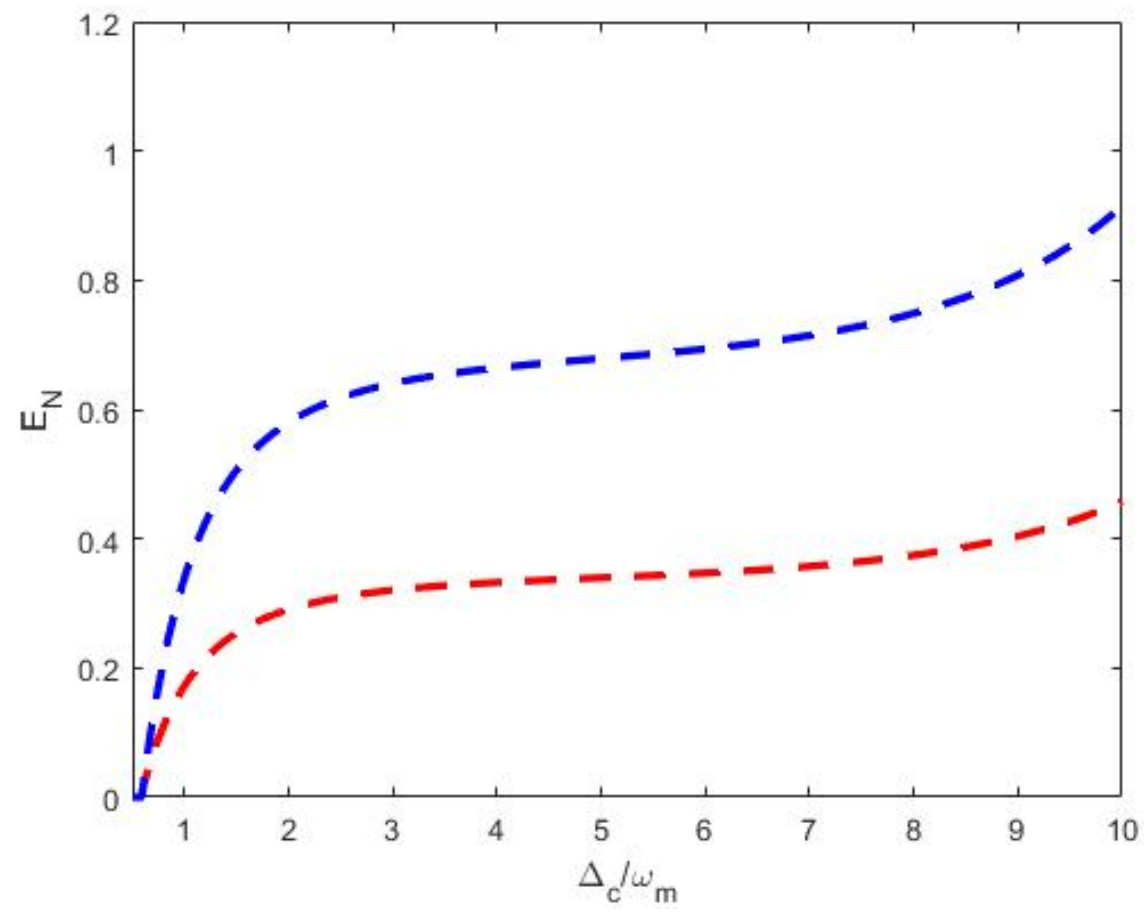

Figure 3: The plot of logarithmic negativity $E_N$ between MC-MR versus normalize optical cavity detuning $\Delta_c/\omega_m$, without injection of atom (red line) and with injection of atom (blue line), the parameters we used are $\kappa_c = 0.08\omega_m$, $r_a = 2000$, $\Delta_{a1} = \Delta_{a2} = 2\pi\times10^7$, $\gamma_m = \omega_m/Q$, $g = 2\pi\times1.0\times10^5$, $\kappa_a = 2\pi\times10^6$, $\Delta_w = \omega_m$ and the others parameters are the same as Fig. (2).

Fig. (4) illustrates an entanglement setup involving OC-MC subsystems. In this case, the red dashed line represents the entanglement between the optical and microwave cavities. In contrast, the blue dashed line signifies the entanglement between OC-MC, with the three-level atom being injected into the optical cavity at a rate of $r_a$. It is essential to highlight that the entanglement between OC-MC is not achieved through direct interaction but is mediated by the mechanical resonator.

On the other hand, the blue dashed line represents the entanglement between OC-MC when the three-level laser is present. As the atom is injected into the OC, the emitted laser experiences a coupling between the optical cavity field mode. This coupling induces an indirect interaction between the OC-MC, promoting entanglement generation. When we compare the two situations, it is clear that the presence of the three-level laser significantly improves the entanglement between the optical and microwave cavities. This suggests that the injection of the atom alters the dynamics of the OC, MC, and MR system, which leads to the enhancement of the entanglement process. Moreover, by controlling the injection rate, the level of entanglement can be adjusted according to desired parameters.

The inclusion of a three-level laser can also increase the system's coherence and stability, which in turn can improve performance in a range of quantum applications like quantum information processing and communication, where robust entanglement between quantum systems is desirable. In general, in the above three bipartite entanglement between subsystems OC-MR, OC-MC, and MR-MC, it is shown that the subsystem containing OC has higher logarithmic negativity due to the atom injected into the optomechanical arrangement and emitted laser, which interact with OC. Also, it is shown that the entanglement between subsystems of the opto-electro-mechanical system discussed above is indeed enhanced compared to the entanglement of opto-electro-mechanical studied by [42, 36].

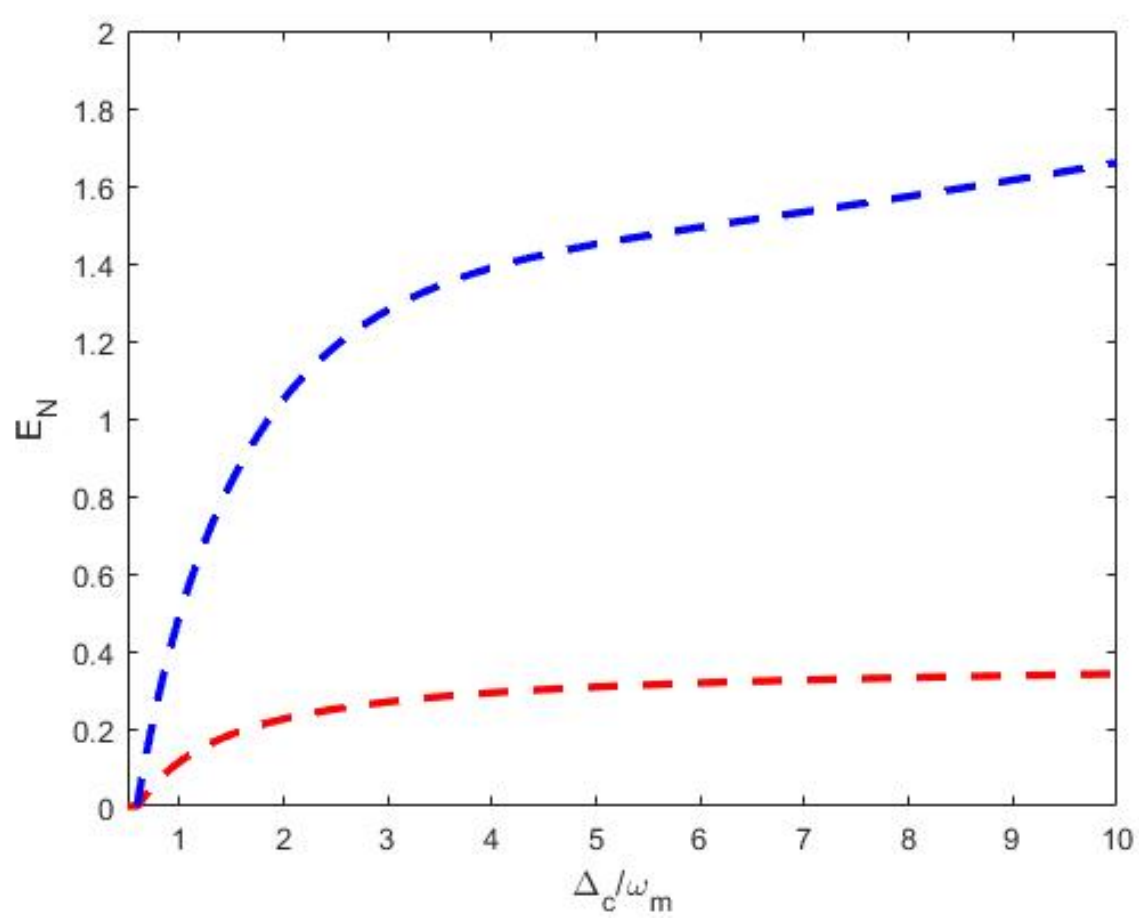

Figure 4: The plot of logarithmic negativity $E_N$ between OC-MC versus normalize optical cavity detuning $\Delta_c/\omega_m$, without injection of atom (red line) and with the injection of atom (blue line), the parameters we used are $r_a = 1.6\times10^6$, $\Delta_{a1} = 2\pi\times10^{10}$, $\Delta_{a2} = 2\pi\times10^6$, $g = 2\pi\times1.5\times10^6$, $\kappa_c = 0.08\omega_m$ and the others parameters are the same as Fig. (2).

Fig. (5) is a plot of logarithmic negativity $E_N$ between OC-three level laser against normalized optical cavity detuning $\Delta_c/\kappa_c$ for different three-level laser-cavity field coupling rates $g$. Three scenarios are presented in the plot, distinguished by different colored solid lines. The purpose of these three scenarios is to demonstrate how the entanglement between the optical cavity and three level laser varies as the coupling rate is increased. As the coupling rate increases, represented by the color change from red to blue, the entanglement (logarithmic negativity) also increases. This implies that the degree of correlation between the two subsystems, the optical cavity, and three-level lasers, becomes stronger as the three-level laser-cavity field coupling rate is increased [69]. Furthermore, the dependence of logarithmic negativity on the normalized optical cavity detuning ($\Delta_c/\kappa_c$) can be observed as well. The entanglement increases as the detuning is increased but reaches a peak at a particular value of $\Delta_c/\kappa_c$ and then starts to decrease. This shows an optimal detuning value that maximizes the entanglement between the optical cavity and the three-level laser for a given coupling rate. The peak entanglement value, i.e., the maximum logarithmic negativity and the optimal detuning value at which this maximum entanglement occurs, increases as the coupling rate increases.

As shown in Fig. (6), the logarithmic negativity $E_N$ is plotted is plotted for three bipartite entanglements against the normalized optical cavity detuning $\Delta_c/\omega_m$ at three different temperatures. The first sub-plot (a) illustrates the logarithmic negativity at a temperature of $5mK$. It is evident that the entanglement between the optical OC-MC is higher compared to the other subsystems. The second sub-plot (b) shows the behavior at $250mK$. Here, we observe that the entanglement between each subsystem decreases as the temperature increases. Finally, the third sub-plot (c) presents the logarithmic negativity at $350mK$. The trend shows a further reduction in entanglement as the temperature continues to rise.

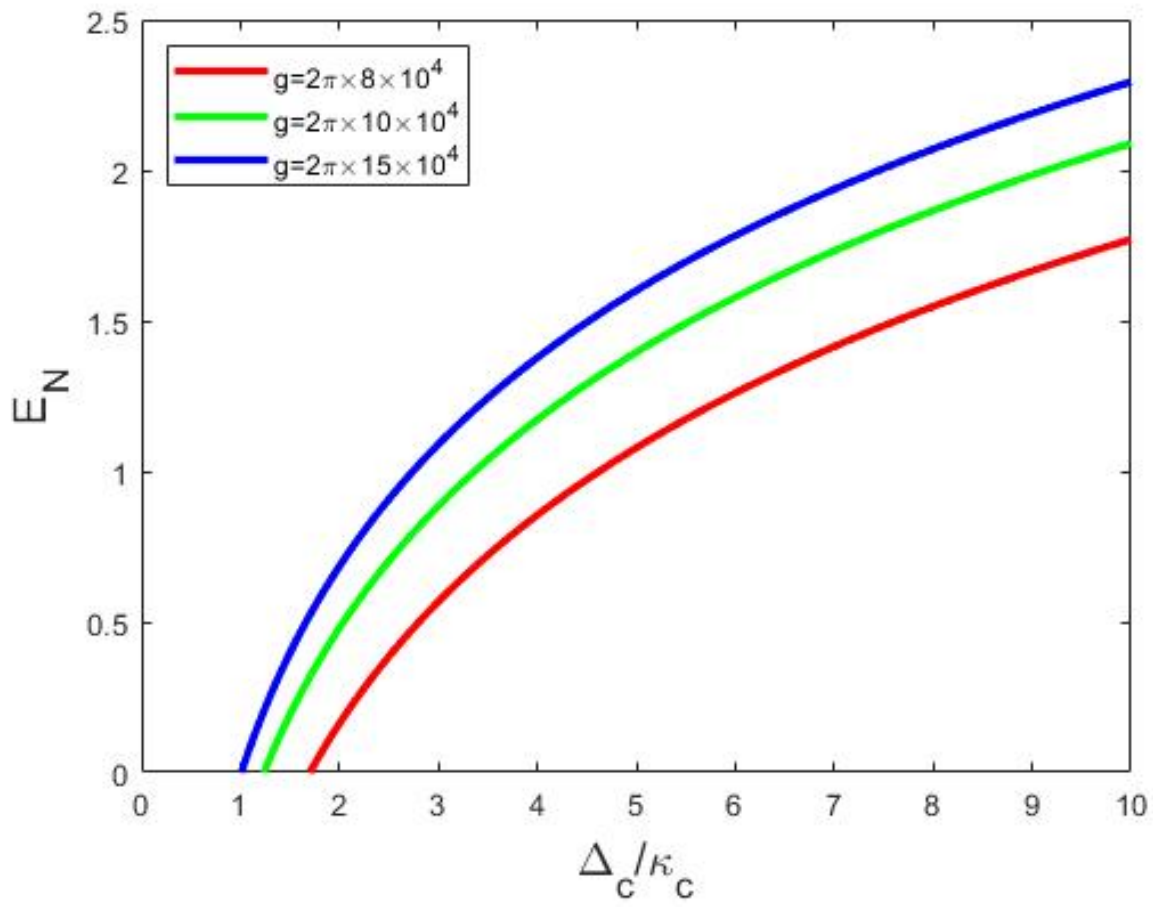

Figure 5: The plot of logarithmic negativity $E_N$ between OC-three level laser versus normalize optical cavity detuning $\Delta_c/\kappa_c$ at different three-level laser-cavity field coupling rate, the parameters we used are $r_a = 1.6 \times 10^6$, $\Delta_{a1} = \Delta_{a2} = 2\pi \times 10^6$, $\kappa_a = 2\pi \times 10^5$, $\kappa_c = 0.02$ and the others parameters are the same as Fig. (2).

Different colored lines represent the entanglement between the various subsystems. Specifically, it can be observed that the entanglement between each subsystem decreases as the temperature increases from $5mK$ (a) to $350mK$ (c). At temperatures above $450mK$, the thermal fluctuations and decoherence processes will lead to the sudden disappearance of the bipartite entanglement. The entanglement sudden death at the temperature of 450 $mK$, suggests that the system has a relatively strong coupling between subsystems, as stronger coupling generally leads to higher entanglement sudden death temperatures. This can be attributed to the fact that entanglement is a quantum mechanical property that tends to become fragile with increasing temperature [61]. As the environment becomes warmer, the increased thermal fluctuations lead to greater decoherence and noise, negatively impacting the entanglement between the various subsystems.

Moreover, by adjusting system parameters, it is possible to control the level of entanglement, offering potential for designing robust entangled systems with practical applications. This tunability also suggests that the entanglement between subsystem pairs is highly sensitive to temperature, with higher temperatures accelerating entanglement decay.

Comparing the peaks of the plotted lines at different temperatures reveals that the entanglement between the OC-MC pair decays more significantly at higher temperatures compared to the entanglement in the OC-MR and MC-MR pairs. This can be explained by the fact that the entanglement between OC-MC is enhanced by the entanglement in the OC-MR and MC-MR pairs, making it more susceptible to temperature-induced decay.

Moreover, by altering the system parameters, one can control the level of entanglement. This entanglement tunability may have implications for designing robust entangled systems for various practical applications. This relationship also suggest that the behavior of the entangled subsystem pairs is highly sensitive to temperature, with higher temperatures leading to decay entanglement between the subsystems. When we compare the peaks of the plotted lines at different temperatures, it is evident that the entanglement between the OC-MC pair significantly decays at higher temperatures than the entanglement in the OC-MR and MC-MR pairs. This is because the entanglement between OC-MC is enhanced by the entanglement in the OC-MR and MC-MR pairs, making it more susceptible to temperature-induced decay.

In Fig.(7), we plot the tripartite entanglement as a function of the normalized optical cavity detuning $\Delta_c/\omega_m$ for different values of atomic decay rate $\kappa_a$. The plot in Fig. (7) shows that tripartite entanglement is sensitive to both the optical cavity detuning and the atomic decay rate. At lower values of $\kappa_a$, the system can maintain higher degrees of entanglement, especially at certain optimal values of the detuning $\Delta_c/\omega_m$. This suggests a balance point where the detuning compensates for the energy dissipation to maximize entanglement. As $\kappa_a$ increases, the coherence necessary for maintaining entanglement is gradually lost, leading to a monotonic decrease in entanglement. The physical reason is that higher decay rates mean that the atoms interact more strongly with their environment, which tends to "measure" the quantum states, thus destroying the entanglement. The rapid energy dissipation also means that the system is less able to maintain the correlations needed for entanglement. Furthermore, the rise and subsequent decline in entanglement suggest that there exists an optimal detuning that maximizes entanglement for a given decay rate. Beyond this optimal point, further detuning likely pushes the system out of the resonant conditions necessary to maintain strong quantum correlations, reducing entanglement.

## 5. Conclusion

In conclusion, the study focused on improving bipartite entanglement in an opto-electro-mechanical system utilizing three-level atoms. A system's dynamics have been studied using the QLEs and linearization approximation, leading to a correlation matrix containing block matrices describing entanglement between each subsystem. Accordingly, the bipartite entanglement is quantified through logarithmic negativity. Specifically, we investigate the entanglement properties generated between various subsystems, including the OC-MC, OC-MR, and MC-MR subsystems. The numerical results indicated that introducing three-level atoms into opto-electro-mechanical systems significantly enhances the entanglement between each subsystem. In addition, we showed that the degree of correlation between the optical cavity and three-level lasers becomes stronger as the three-level laser-cavity field coupling rate increases. A higher coupling rate allows for faster and more reliable information exchange, enabling efficient communication and synchronization between different system parts. Moreover, the results demonstrated that temperature has a significant impact on entanglement, with higher temperatures leading to a reduction in entanglement between the subsystems. This increased interaction leads to a decrease in entanglement between the subsystems as their states become more independent due to the influence of the environment. , the hybrid electro-opto-mechanics system provides a framework for information transfer through tripartite entanglement quantifiers and allows for a detailed examination of how tripartite entanglement responds to variations in atomic decay rates.

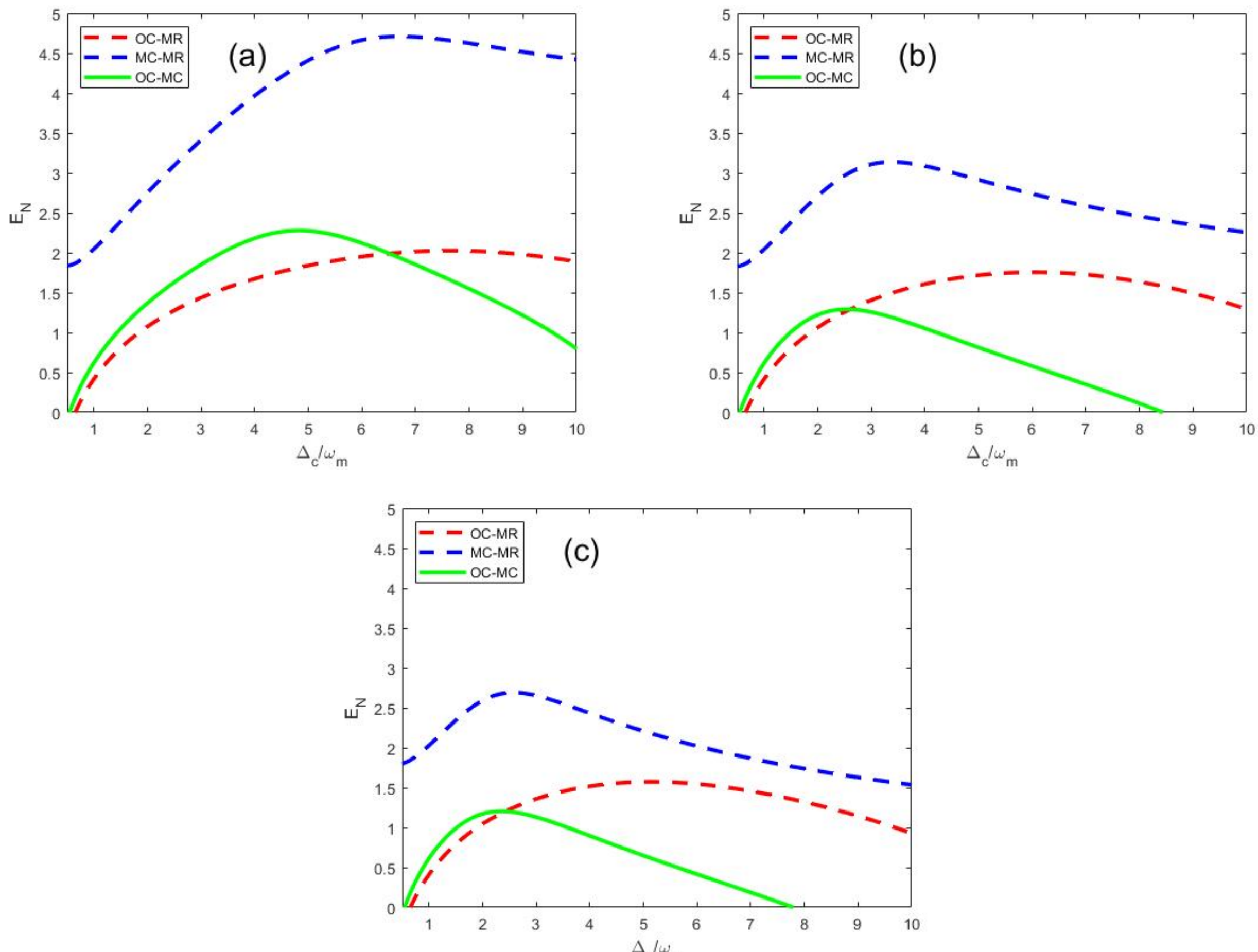

Figure 6: The plot of $E_N$ of the three bipartite sub-systems versus the normalized optical cavity detuning $\Delta_c/\omega_m$ at three different temperatures: $T = 5mK$ (a), $T = 250mK(b)$,T=350mK$(c), theparametersweusedarer_a = 1.6 \times 10^6$, $\Delta_{a1} = 2\pi \times 10^{10}$, $\Delta_{a2} = 2\pi \times 10^6$, $\gamma_m = 200\pi$, $F = 4.07 \times 10^4$, $\kappa_c = \pi c/(F \times L)$, $g = 2\pi \times 1.0 \times 10^5$, $\kappa_a = 2\pi \times 10^6$, $\Delta_w = -\omega_m$ and the others parameters are the same as Fig. (2).

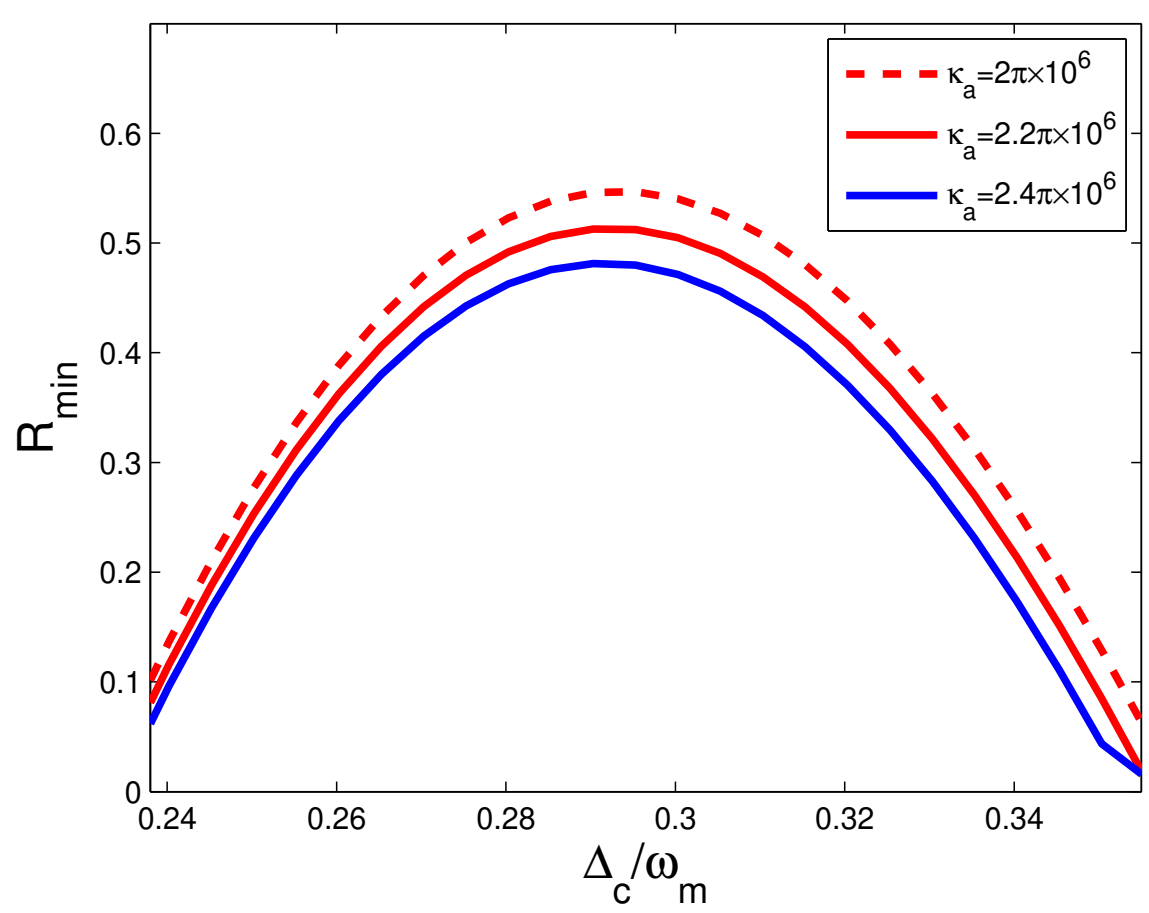

Figure 7: The plot of tripartite entanglement $R_{min}$ versus normalize optical cavity detuning $\Delta_c/\omega_m$ at different values of atomic decay rate $\kappa_a$, the parameters we used are $r_a = 1.2\omega_m$, $\Delta_{a1} = \Delta_{a2} = \omega_m$, $\kappa_w = 0.5 \times \omega_m$, $g = 2.5\pi \times 10^5$, $T = 5mK$ and the others parameters are the same as Fig. (2).

## CRediT authorship contribution statement

**Abebe Senbeto Kussia**: Mathematical analysis, Numerical simulation. **Abeba Teklie**: Analyzing the physics model. **Tewodros Y**: Visualization, Validation. **Berihu Teklu**: Writing, editing, and reviewing, **Tesfay Gebremariam**: Supervision, Methodology, Writing, and reviewing.

### Disclosures.

The authors declare that they have no known competing financial interests or personal relationships that could have appeared to influence the work reported in this paper.

### Data availability.

All data generated and analyzed are presented in the research presented.

### Acknowledgment.

The authors thank to Mr. Habtamu Dagnaw Mekonnen for the useful discussions. We gratefully acknowledge, Arba Minch University, Salale University, and Adama Science and Technology University for their support during this work.